\documentclass[aip,jcp,reprint,twocolumn,notitlepage,eqsecnum,floatfix]{revtex4-2}

\usepackage{amsmath}
\usepackage{color}
\usepackage{graphicx}
\usepackage{hyperref}
\usepackage{multirow}

\newcommand{\bq}{\begin{eqnarray}}
\newcommand{\eq}{\end{eqnarray}}
\newcommand{\bqn}{\begin{eqnarray*}}
\newcommand{\eqn}{\end{eqnarray*}}
\newcommand{\bqs}{\begin{subequations}}
\newcommand{\eqs}{\end{subequations}}
\newcommand{\bw}{\begin{widetext}}
\newcommand{\ew}{\end{widetext}}

\newcommand{\kk}{{\bf k}}
\newcommand{\rr}{{\bf r}}
\newcommand{\qq}{{\bf q}}
\newcommand{\QQ}{{\bf Q}}

\newcommand{\calo}{{\cal O}}
\newcommand{\nobarfrac}{\genfrac{}{}{0pt}{}}

\newcommand{\red}[1]{{#1}}

\begin{document}
\title{Monte Carlo simulation of Hard-, Square-Well, and Square-Shoulder Disks in 
narrow channels}

\author{Riccardo Fantoni}
\email{riccardo.fantoni@scuola.istruzione.it}
\affiliation{Universit\`a di Trieste, Dipartimento di Fisica, strada
  Costiera 11, 34151 Grignano (Trieste), Italy}

\date{\today}

\begin{abstract}
We perform Monte Carlo simulation of the thermodynamic and structural properties of 
Hard-, Square-Well, and Square-Shoulder Disks in narrow channels. For the 
thermodynamics we study the internal energy per particle and the longitudinal and 
transverse compressibility factor. For the structure we study the Transverse Density 
and Density of Pairs Profiles, the Radial Distribution Function and Longitudinal 
Distribution Function, and the (static) Longitudinal Structure Factor. We compare 
our results with a recent exact semi-analytic solution found by Montero and 
Santos for the single file formation and first nearest neighbor fluid and 
explore how their solution performs when these conditions are not fulfilled making 
it just an approximation.
\end{abstract}


\maketitle
\section{Introduction}

{\sl Confined fluids} are an important field of study
due to the wide range of applications and situations
where they can be found \red{\cite{Fantoni13d}}.
Interesting systems in physics, chemistry, or biology involve dealing with 
confined particles. Examples are carbon nanotubes \cite{Kuakuno2011,Majumder2011} 
or biological ionchannels \cite{Boda2008}. In many of these systems, the 
geometry is so restrictive that one or more spatial dimensions become negligible. 
One can therefore often describe these systems as living in a one (1D) or two (2D) 
dimensional space in order to simplify the mathematical model and its subsequent 
study. Yet, in some cases it is necessary a more realistic description which
can be obtained by modeling the geometrical restriction without recurring to 
a dimensionality reduction. So for particles living in three dimensions we will 
talk about {\sl quasi} two dimensional (quasi 2D) or {\sl quasi} one 
dimensional (quasi 1D) fluids. In this work we will study particles living
in 2D which are quasi 1D.
 
Despite its clear importance, systems whose structural properties are amenable to
exact analytic solutions are very scarce, and usually limited to 
1D fluids \cite{Fantoni16d, Fantoni17a, Fantoni21a, Fantoni08b, Fantoni10a, 
Fantoni10b, Fantoni13c, Fantoni13h, Fantoni21g, Fantoni17e} with only nearest 
neighbor interactions \cite{Santosb, Tonks1936, Salsburg1953}. 
Even if for restricted values of the thermodynamic parameters even 2D 
fluids may offer an exact analytic classical equilibrium statistical mechanics 
solution \cite{Fantoni03a,Fantoni08c,Fantoni12b,Fantoni21i}. Otherwise, one must 
resort to approximations, numerical methods, or simulations.

Recently, Montero and Santos \cite{Montero2023a,Montero2023} 
developed an exact semi-analytic formalism able to solve the {\sl longitudinal} 
structure and thermodynamics of a quasi 1D problem 
of {\sl single file} formation and {\sl first nearest neighbor} 
fluids of {\sl hard\red{-core}} particles in {\sl narrow channels}.

In particular the single file confinement constraint \cite{Poncet2021,Horner2018}
implies that particles are inside a {\sl pore} that is not wide enough to allow 
particles to bypass each other, therefore confining them into a single file 
formation.

Whereas the first nearest neighbor constraint implies that the particles are 
not allowed to interact with their second (or beyond) nearest neighbors 
\cite{Fantoni17e}. 

The pore is a 2D narrow channel or band with {\sl periodic 
boundary conditions} along the longitudinal direction and {\sl open boundary 
conditions} along the transverse direction where the particles are assumed to 
be confined by {\sl hard walls}.

The nearest neighbor constraint allows the use of the exact solution that is 
available for 1D fluids subject to such 
constraint\cite{Tonks1936,Salsburg1953,Santosb}. 
In fact Montero and Santos study their quasi 1D fluids of particles 
interacting through a {\sl pairwise} potential $\varphi_{2D}(r)$ with a mapping 
to a 1D {\sl non-additive} mixture \red{\cite{Fantoni11e}} of equal chemical potentials species, where the 
species index $i$ denotes those particles with the ordinate equal to a fixed 
value within the channel and the interaction potential becomes 
$\varphi_{ij}(x)=\varphi_{2D}(\sqrt{x^2+(y_i-y_j)^2})$. They further 
assume the mixture to be {\sl polydisperse} \cite{Fantoni05a,Fantoni06b,Fantoni06c}
so that the molar fraction $x_i$ of 
the $i$th species can be rewritten as $F(y)dy$ which represents the fraction of 
particles with the ordinate lying in the interval $[y,y+dy]$. What they find 
\cite{Kofke1993} is that working in the isothermal isobaric ensemble the average 
of a function of $y$ can be expressed as $\langle f(y)\rangle=\int dy\,F(y)f(y)$ 
where $F(y)=\phi^2(y)$\red{. Here} $\phi(y)$ is the eigenfunction of the maximum 
eigenvalue of a certain operator 
$K(y_1,y_2)=\exp\{-\beta P_L\sigma(y_1,y_2)-\frac{1}{2}\beta[\Phi_{\rm ext}(y_1)+\Phi_{\rm ext}(y_2)]\}$ where $\beta=1/k_BT$ with $k_B$ Boltzmann constant and 
$T$ the absolute temperature, $P_L$ is the longitudinal pressure, 
$\sigma(y_1,y_2)$ is the distance of closest approach of particles 1 and 2, 
which are nearest neighbors, and $\Phi_{\rm ext}(y)$ is the external potential 
which acts only in the dimension of the $y$ coordinates and confines the particles 
within the channel.

In this work we will perform Monte Carlo (MC) simulations 
\cite{Mon2020,Kofke1993b,Varga2011}, in the {\sl canonical
ensemble} (for results using molecular dynamics see Ref. 
\cite{Huerta2021}), to study the thermodynamic and structural properties 
of Hard (HD), Square-Well (SW), and Square-Shoulder (SS) disks in narrow channels.
For the thermodynamics we will study the internal energy and the compressibility 
factor. For the structure we will study the Transverse Density Profile (TDP), the
Transverse Density of Pairs Profile (TDPP), the Radial Distribution Function (RDF) 
and Longitudinal Distribution Function (LDF), and the (static) Longitudinal 
Structure Factor (LSF). 

Of course with our MC simulations we are not bound to fulfill the single file
and nearest neighbor constraints. We will therefore also study the performance of 
the solution of Montero and Santos outside the nearest neighbor regime where 
it is expected to be just an approximation.

The work is organized as follows: In Section \ref{sec:model} we will describe the 
mathematical model of the physical fluid of interest, the MC estimators
for the quantities we want to measure in our computer experiments, and some MC 
results for the thermodynamics. In Section \ref{sec:results} we present our MC 
results for the structure. Section \ref{sec:conclusions} is for concluding remarks.
 
\section{Model and simulation details}
\label{sec:model}
Consider a 2D system of $N$ particles interacting
via a {\sl pairwise potential} $\varphi_{\rm 2D}(r)$. The particles
are confined in a very long channel of width $w=1+\epsilon$ and
length $L\gg w$, in such a way that they are in {\sl single file}
formation and only {\sl first nearest neighbor} interactions
take place. The {\sl channel surface density} of the fluid will be 
$\sigma=N/Lw=\lambda/w$ with $\lambda$ the {\sl longitudinal density}. 
The total potential energy of the fluid will be
\bq
\Phi(\QQ)=\frac{1}{2}\sum_{i,i\neq j}\varphi_{\rm 2D}(q_{ij})+\Phi_{ext},
\eq
where $\QQ=(\qq_1,\qq_2,\ldots,\qq_N)$ are the positions of the particles 
in the channel and $\qq=(x,y)$ with $x\in[-L/2,L/2]$ and 
$y\in[-\epsilon/2,\epsilon/2]$. We have {\sl periodic boundary conditions} 
(PBC) along $x$ so that $x_i\to x_i-{\rm nint}(x_i/L)L$ where `nint' is the 
nearest integer and enforce the usual minimum image convention so that 
$x_i-x_j\to x_i-x_j-{\rm nint}[(x_i-x_j)/L]L$. Along $y$, instead, we have 
{\sl open boundary conditions} (OPC) where in particular we assume to have 
an infinitely repulsive {\sl external potential} $\Phi_{ext}$ for 
$y>\epsilon/2$ and $y<-\epsilon/2$ and we don't employ the minimum image 
convention. We will denote with $q_{ij}=\sqrt{(x_i-x_j)^2+(y_i-y_j)^2}$ the 
distance between particles at $\qq_i$ and at $\qq_j$. 

For Hard Disks (HD) we have
\bq \label{pp-hd}
\varphi_{\rm 2D}(r)=\left\{\begin{array}{ll}
\infty & \mbox{if $r<1$}\\
0      & \mbox{else}
\end{array}\right..
\eq
If the transverse separation between two disks at contact is $s$, their 
longitudinal separation is
\bq
a(s)=\sqrt{1-s^2}.
\eq
The single file constraint in this case requires clearly 
$\epsilon<\epsilon_{\rm sf}=1$. In this case we have a close packing 
limit longitudinal density given by $\lambda_{\rm cp}=1/a(\epsilon)$. In order to 
enforce also the first nearest neighbor constraint we require 
$\epsilon<\epsilon_{\rm nn-HD}=\sqrt{3}/2$. For $\epsilon=\sqrt{3}/2$ 
the close packing longitudinal density is $\lambda_{\rm cp}=2$ and the surface 
density is $\sigma_{\rm cp}=2/(1+\sqrt{3}/2)=1.071\ldots$. 

For Square-Wells (SW) or Square-Shoulders (SS) we have
\bq \label{pp-sw}
\varphi_{\rm 2D}(r)=\left\{\begin{array}{ll}
\infty     & \mbox{if $r<1$}\\
-\varphi_0 & \mbox{if $1<r<r_0$}\\
0          & \mbox{else}
\end{array}\right.,
\eq
with $\varphi_0>0$ for SW and $\varphi_0<0$ for SS.

In this case in order to enforce the first nearest neighbor constraint we require
$\epsilon<\epsilon_{\rm nn}=\sqrt{1-(r_0/2)^2}$. Since 
$r_0>1$ we will have $\epsilon_{\rm nn}<\epsilon_{\rm nn-HD}$.

In our computer experiment we measured various thermodynamic and structural 
properties of these fluids. We could than compare our numerical {\sl meta data} 
with analytic or semi-analytic theoretical data available in the literature.
To {\sl measure} an observable $\calo$ we need to calculate \cite{Kalos-Whitlock} 
the following quantity
\bq
\langle\calo\rangle=\frac{\int
  O(\QQ)\exp[-\beta\Phi(\QQ)]\,d\QQ}
{\int\exp[-\beta\Phi(\QQ)]\,d\QQ},
\eq
where $\beta=1/k_BT$ with $k_B$ Boltzmann constant and $T$ absolute temperature. In 
our canonical (at fixed number of particles, surface area, and temperature) Monte
Carlo (MC) simulation we employed the usual M(RT)$^2$ algorithm 
\cite{Kalos-Whitlock} to sample the probability distribution 
$\propto\exp[-\beta\Phi(\QQ)]$.

We generally found it sufficient to use $N=100$ with runs up to $10^9$ MC single
particle moves long. The spatial extent of the uniform particle displacement move
was tuned so to have acceptance ratios around $1/2$ and kept constant during the 
run, even if this was not always possible at high densities.

\subsection{Structure}
For the {\sl Radial Distribution Function} (RDF) 
\cite{Allen-Tildesley,Frenkel-Smit}, $g(r)=\langle\calo\rangle$, we have the
following {\sl histogram estimator}
\bq \label{rdf}
O(\QQ;r)=\sum_{i,i\neq j}
\frac{1_{[r-\Delta/2,r+\Delta/2[}\,(q_{ij})}{Nn_{\rm id}(r)},
\eq
where $\Delta$ is the histogram bin, $1_{[a,b[}(t)=1$ if $t\in[a,b[$ and 0
otherwise, and $n_{\rm id}(r)$ is the average number of particles on the 
interception of the circular crown $[r-\Delta/2,r+\Delta/2[$ with the 
part of the channel accessible to the particles centers, for the uniform gas at 
the same longitudinal density $\lambda$, which for a narrow channel can be 
approximated to
\bq
n_{\rm id}=2\lambda\Delta~,
\eq
independent of $r$. We have that $g(r)$ gives the 
probability density that sitting on a particle at $\qq$ one has to find another 
particle at $\qq^\prime$, where $r=\sqrt{(x-x')^2+(y-y')^2}$.  

Instead of counting how many disks are separated a 2D distance $r$ we can count 
how many are separated a 1D longitudinal distance $|x|$. So, doing the same
calculation described above, but, in the end, keeping track only of the relative 
abscissas, $r=\sqrt{(x-x')^2}=|x-x'|$, of the particles, we find the quasi 1D or 
{\sl Longitudinal Distribution Function} (LDF) $g(|x|)$.

The Fourier transform of the radial distribution function is the (static) 
structure factor $S(k)$ which for an isotropic system is given by
\bq \nonumber 
S(k)&=&1+\frac{\lambda}{\epsilon}\int[g(r)-1]\exp(-i\kk\cdot\rr)\,d\rr\\ 
\label{sk}
&&+\frac{\lambda}{\epsilon}(2\pi)^2\delta(\kk),
\eq
where usually the Dirac delta function is neglected. Note also that from the
definition (\ref{rdf}) we find the following sum rule
\bq
\frac{\lambda}{\epsilon}\int[g(r)-1]\,d\rr=-1,
\eq
and from the definition (\ref{sk}) follows $\lim_{k\to 0}S(k)=0$. 
Moreover if $\lim_{r\to 0}g(r)=0$ then $\lim_{k\to\infty}S(k)=1$.
Now for our quasi 1D geometry of the narrow channel, the isotropy 
is clearly lost and when we count just the longitudinal distances between the 
particles, for the LDF $g(|x|)$, we may still find $\lim_{k_x\to 0}S(k_x)\neq 0$ 
since the sum rule becomes 
$\frac{\lambda}{\epsilon}\int[g(|x|)-1]\,dx=-\frac{1}{\epsilon}<-1$. 
In the following we will always refer to this {\sl Longitudinal Structure 
Factor} (LSF) $S(k_x)$ and for brevity we will simply rewrite $k_x\to k$.

Another structural property to study is the {\sl Transverse Density Profile} (TDP)
$F(y)$ such that $F(y)dy$ gives the fraction of particles with the ordinate in the 
interval $[y,y+dy]$. By symmetry we clearly must have for $F$ an even function.
A related function is $F_2(y)$, the fraction of pairs of different particles 
1 and 2 such that their transverse distance $|y_2-y_1|\in[y,y+dy]$. We will call 
this the {\sl Transverse Density of Pairs Profile} (TDPP).

\subsection{Internal energy}
For the {\sl internal energy per particle} of the fluid 
\cite{Allen-Tildesley,Frenkel-Smit} we have $u=\langle\calo\rangle$ with the 
following internal energy per particle {\sl estimator}
\bq \label{eq:iepp}
O(\QQ)&=&\Phi(\QQ)/N. 
\eq

For SW/SS with $|\beta\varphi_0|=1$ we found the results of Table 
\ref{tab:tabI}.

\bw
\begin{center}
\begin{table}[h!]
\caption{Internal energy per particle (\ref{eq:iepp}) and total pressure 
(\ref{eq:Z2DE}) for $N=100$ SW/SS with 
$|\beta\varphi_0|=1$. The results were determined from runs made of 
$5\times 10^7$ single particle moves.}
\label{tab:tabI}
\begin{center}
\begin{tabular}{||c|c|c|c|c||c|c||c|c||}
\hline
\hline
$\epsilon$ & $r_0$ & $\lambda_{\rm cp}$ & $\lambda$ & $\sigma$ & \multicolumn{2}{ |c|| }{$u$} & \multicolumn{2}{ |c|| }{$\beta P$} \\ 
\hline
\multicolumn{5}{ ||c|| }{} & SW & SS & SW & SS \\ 
\hline
\hline
4/5          & 6/5 & 1.66667 & 1.080 & 0.6 & $-$0.9303(4)    & +0.7673(7) & 7.351(1) & 9.965(3) \\
4/5          & 6/5 & 1.66667 & 1.260 & 0.7 & $-$0.9797(4)    & +0.9176(8) & 14.653(4) & 16.45(1) \\
$\sqrt{7}/4$ & 3/2 & 1.33333 & 0.997 & 0.6 & $-$0.99837(2)   & +0.9895(2) & 9.865(2) & 10.160(2)\\
$\sqrt{7}/4$ & 3/2 & 1.33333 & 1.163 & 0.7 & $-$0.9999989(6) & +0.99998(1)& 23.856(7) & 23.12(2) \\
\hline
\hline
\end{tabular}
\end{center}
\end{table}
\end{center}
\ew

\subsection{Compressibility factor}
For the compressibility factor $Z=\beta P/(\lambda/\epsilon)$ of the confined
2D fluid we have, from the virial theorem \cite{Hansen-McDonald} 
\bq \nonumber
Z&=&1-\beta\frac{\lambda}{\epsilon}\frac{1}{4}\int\!\!\!
\int_{\nobarfrac{x_1, x_2\in[-L/2,L/2]} 
{y_1, y_2\in[-\epsilon/2,\epsilon/2]}}
r\varphi_{\rm2D}'(r)g(\rr)\,d\rr\\
\nonumber
&\approx &1+\frac{\lambda}{2}\int_0^\infty
\frac{d\exp[-\beta\varphi_{\rm 2D}(r)]}{dr}ry(r)\,dr\\ \label{eq:Z2DE}
&=&1+\frac{\lambda}{2}[g(1^+)+
(1-e^{\beta\varphi_0})r_0g(r_0^+)],
\eq
where in the first line $d\rr=d(x_1-x_2)d(y_1-y_2)$, 
$g(\rr)=g(|x_1-x_2|;y_1,y_2)$, and we used polar coordinates so that 
$d\rr=rd\theta dr$. In the second line we approximated 
$\int_{\rm channel}\,rd\theta\approx 2\epsilon$ for the narrow channel, we 
then introduced the continuous indirect correlation function 
$y(r)=g(r)\exp[-\beta\varphi_{\rm 2D}(r)]$, where $g(r)$ is the 2D RDF of 
Eq. (\ref{rdf}), and used the fact that for the SW/SS pair potential of Eq. 
(\ref{pp-sw}) we have
\bq \nonumber
\frac{d\exp[-\beta\varphi_{\rm 2D}(r)]}{dr}&=&e^{\beta\varphi_0}\delta(r-1)\\
&&+(1-e^{\beta\varphi_0})\delta(r-r_0).
\eq

The total thermodynamic pressure $P=(P_L+P_T)/2$, where $P_L$ and $P_T$ are the 
{\sl longitudinal and transverse 2D pressures}, respectively. In Table 
$\ref{tab:tabI}$ we present some results for SW/SS.

Let us now specialize to the HD case so that $\varphi_0=0$. We will also 
introduce $p=P_L\epsilon$. From the Table I of Ref. \cite{Montero2023a} we find 
that $Z_L^{\rm exact}=\beta p/\lambda=12.774$ when $\epsilon=0.4$ and $\beta p=12$. 
From these data we extract $\lambda=\beta p/Z_L^{\rm exact}=0.671 w$ and, at 
this longitudinal density, our canonical simulation gives $\beta P=15.85(2)$ for 
$\Delta=10^{-2}$ and $\beta P=16.53(1)$ for $\Delta=10^{-3}$. Since the exact
longitudinal pressure is $\beta P_L^{\rm exact}=\beta p/\epsilon=30$, we estimate 
a transverse pressure of $\beta P_T\approx 2(16.53)-30=3.06$. For the case when 
$\epsilon=0.8$ and $\beta p=12$ we find $\beta P=14.46(1)$ (these measures tend 
to slightly increase even further at lower $\Delta$). See Table 
\ref{tab:tabII} for these $Z$ measurement. Exact results from the Montero and Santos 
analysis \cite{Ana-P} give $\beta P=16.722$ for $\epsilon=0.4, \beta p=12$ and 
$\beta P=14.8551$ for $\epsilon=0.8, \beta p=12$. \red{Alternatively one can use the 
canonical ensemble exact expression found by Pergamenshchik \cite{Pergamenshchik2020}.}

Alternatively one can use the quasi 1D scenario and the LDF $g(|x|)$ to find the 
longitudinal compressibility factor $Z_L=\beta P_L/(\lambda/\epsilon)$. To do this 
we need to calculate (see Appendix \ref{app:B} \cite{Andres-appA})
\bqs
\bq \label{eq:Z1DDD}
Z_L&=&\frac{1-\lambda I_0}{1-\lambda+\lambda^2(I_0-I_1)},\\
I_n&=&\int_{a(\epsilon)}^{1}x^ng(x)\,dx.
\eq
\eqs
We computed the integrals $I_n$ for $n=0,1$ numerically with a 
discretization $\Delta$ on the abscissa $x^i=\Delta i$ with $i=0,1,2,3,\ldots$. In 
Table \ref{tab:tabII} we show our results compared with the ones of Ref. 
\cite{Montero2023a}. Note that for the two high density cases this way of 
estimating numerically $Z_L$ is not useful since for $\lambda\to\lambda_{\rm cp}$
we find 
\bqs
\bq \label{eq:hdlim1}
\lambda I_0 &\to& 1-8\exp\{-\beta p[1-a(\epsilon)]\}=\lambda I_0^{\rm asy},\\
\nonumber
\lambda^2 I_1 &\to& 1-\frac{1}{2}\{1+[2-a(\epsilon)]\lambda\}
(1-\lambda I_0^{\rm asy})\\ \label{eq:hdlim2}
&&=\lambda^2 I_1^{\rm asy},
\eq
\eqs
and both numerator and denominator in Eq. (\ref{eq:Z1DDD}) vanish. In this case 
one can use the analytic expression (see Appendix C of Ref. \cite{Montero2023a})
\bq \label{eq:hp}
Z_L\to 2+a(\epsilon)p=\frac{2}{1-\lambda/\lambda_{\rm cp}},
\eq
valid asymptotically for $\lambda$ near to its close packing limit 
$\lambda_{\rm cp}$.

\begin{figure}[h!]
\begin{center}
\includegraphics[width=9cm]{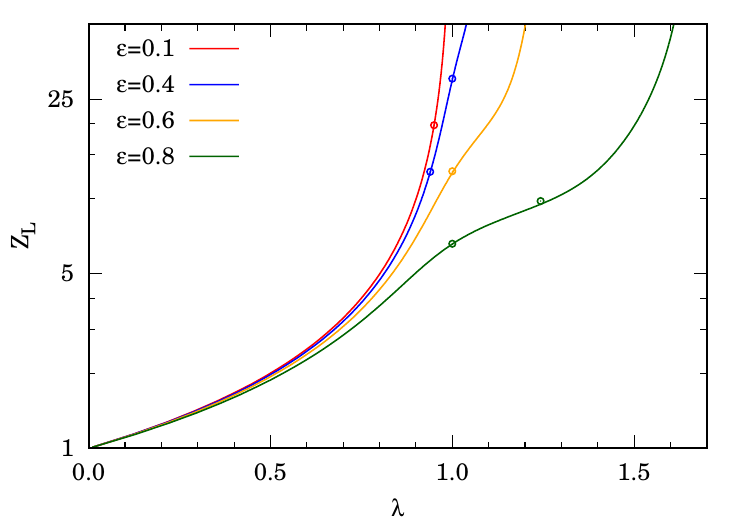}
\end{center}  
\caption{Comparison between our MC (points) and the exact results (lines) of Ref. 
\cite{Montero2023a} for the 
longitudinal compressibility factor. The statistical error in the MC points is 
smaller than the point symbol. The MC simulations were up to $10^9$ single particle
moves long.}
\label{fig:tabI-HD}
\end{figure}

Additional points are presented in Fig. \ref{fig:tabI-HD} where we compare with the 
exact results of Ref. \cite{Montero2023a} (results shared privately and not all
previously published). We found that at the same value of $\lambda$ it takes longer
to equilibrate the large $\epsilon$ cases. For example the MC points at 
$\epsilon=0.1$ required just $10^7$ single particle moves whereas the ones at 
$\epsilon=0.8$ required up to $10^9$ moves. 

Note that a drawback of this way of estimating the longitudinal pressure is that
it is hard to tell if the statistical error is more or less important than the 
systematic error due to the choice of the discretization $\Delta$. In this respect,
instead of working in the canonical $N\lambda T$ ensemble, it would be desirable 
to work in the isothermal isobaric $NpT$ ensemble with a volume change move where 
one only varies the length of the channel $L$.

For the HD case, with $\epsilon=0.4$ and $\lambda=0.671w$, we find the 
TDP $F(y)$ shown in Fig. \ref{fig:tdp-HD}. As you can see the exact result of Ref. 
\cite{Montero2023a} fits our MC data very well. For the same case the TDPP, 
$F_2(y)$, is shown in Fig. \ref{fig:fyy-HD}.
From this figure we can see how the TDPP changes drastically only getting 
really near to the close packing density $\lambda_{\rm cp}=1.091$.

\bw
\begin{center}
\begin{table}[h!]
\caption{Results for $N=100$ HD from Eqs. (\ref{eq:Z2DE}) and (\ref{eq:Z1DDD}) and 
comparison with the exact values of Table I of Ref. \cite{Montero2023a}. \red{$Z_L^{\rm Mon}$ are the MC values of Ref. \cite{Mon2020}}. The two low 
density cases were determined from runs made of $5\times 10^7$ single particle 
moves, $\Delta=10^{-5}$ for $Z_L$, and $\Delta=10^{-3}$ for $Z$.}
\label{tab:tabII}
\begin{center}
\begin{tabular}{||c|c|c|c||c|c|c|c||c|c|c||c|c||}
\hline
\hline
$\epsilon$ & $p$ & $Z_L^{\rm exact}$ & \red{$Z_L^{\rm Mon}$} & $\lambda$ & $\lambda/\epsilon$ & $\lambda_{\rm cp}$ & $2+a(\epsilon)p$ & $Z$ & $Z_L$ & $Z_L^{\rm exact}/Z_L$ & $\lambda(I_0-I_0^{\rm asy})$ & $\lambda^2(I_1-I_1^{\rm asy})$\\ 
\hline
0.4 & 12  & 12.774 & 12.774 & 0.939408 & 2.34852 & 1.09109 & 12.998 & 7.04(1) & 12.8094(3) & 0.997 &2.1&2.3\\
0.4 & 120 & 112.04 & 112.03 & 1.07105 & 2.67762 & 1.09109 & 111.98 & - & ?          & ?     &$+3.6\times 10^{-4}$&$+3.5\times 10^{-4}$\\
0.8 & 12  & 9.6547 & 9.6548 & 1.24292 & 1.55365 & 1.66667 & 9.2000 & 9.31(1) & 9.780(4)   & 0.987 &$-2.4\times 10^{-2}$&$-2.5\times 10^{-2}$\\
0.8 & 120 & 74.017 & 74.016 & 1.62125 & 2.02656 & 1.66667 & 74.000 & - & ?          & ?     &$< 10^{-15}$&$-4.9\times 10^{-6}$\\
\hline
\hline
\end{tabular}
\end{center}
\end{table}
\end{center}
\ew

\section{Results for the structure}
\label{sec:results}
In this Section we present our MC results for the structural properties of the
confined quasi 1D fluids of our interest.

\subsection{Ideal gas (id)}
We first tried to switch off the pair potential between the particles taking 
$\varphi_{2D}(r)=0$ but keeping the confining infinitely repulsive external 
potential $\Phi_{\rm ext}$ switched on. For the case $\lambda=\sigma w=1$ and 
$\epsilon=\sqrt{3}/2$ we found the results for the LDF and RDF shown in Figs.
\ref{fig:gx-id} and \ref{fig:gr-id} respectively. As you can
clearly see from the MC results the LDF, and as a consequence the LSF, is uniform 
but the RDF is not. 

Note that this is just an effect of the geometry of the confinement in fact
using periodic boundary conditions also along the transverse, $y$,
direction one gets both a uniform LDF and RDF as expected. Moreover the TDP
turns out to be unifrom $F(y)=1/\epsilon$ irrespectively of using open or 
periodic boundary conditions along the transverse direction.

For the case of our interest, with periodic boundary conditions along $x$ and open 
boundary conditions along $y$, the RDF can be calculated exactly analytically as 
follows (see Appendix \ref{app:A} \cite{Andres-appA})
\bw
\bq \label{eq:andres}
g_{\rm id}(r)=\frac{2r}{\epsilon}\left\{\begin{array}{ll}
\pi/2-r/\epsilon & r<\epsilon\\
\sqrt{(r/\epsilon)^2-1}-r/\epsilon + \arctan[1/\sqrt{(r/\epsilon)^2-1}]& 
\mbox{else}
\end{array}\right..
\eq

\begin{figure}[h!]
\begin{center}
\includegraphics[width=18cm]{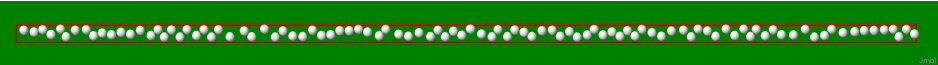}
\end{center}  
\caption{Snapshot of the simulation box for $N=100$ HD of radius 
$r_0=1$ with $\epsilon=\sqrt{3}/2$ and $\lambda=1$.}
\label{fig:snap-HD}
\end{figure}
\begin{figure}[h!]
\begin{center}
\includegraphics[width=18cm]{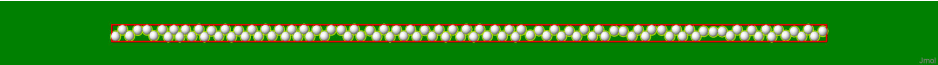}
\end{center}  
\caption{Snapshot of the simulation box for $N=100$ SW with $\beta\varphi_0=1$ 
and radius $r_0=6/5$ with $\epsilon=4/5$ and $\sigma=7/10$.}
\label{fig:snap-SW}
\end{figure}
\ew

\begin{figure}[h!]
\begin{center}
\includegraphics[width=9cm]{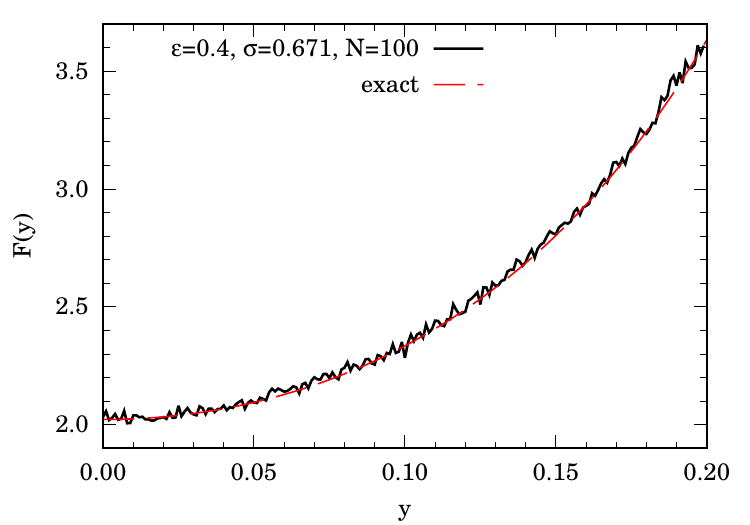}
\end{center}  
\caption{TDP for $N=100$ HD with $\epsilon=0.4$ and $\lambda=0.671w$. 
The exact result, shared privately by A. Montero and not published before,
fits our MC results very well. In particular from Fig. 4 of Ref. 
\cite{Montero2023a} we see how the particles tend to escape from the center of the
channel preferring to stay in contact with the walls as density approaches the
packing density.}
\label{fig:tdp-HD}
\end{figure}
\begin{figure}[h!]
\begin{center}
\includegraphics[width=9cm]{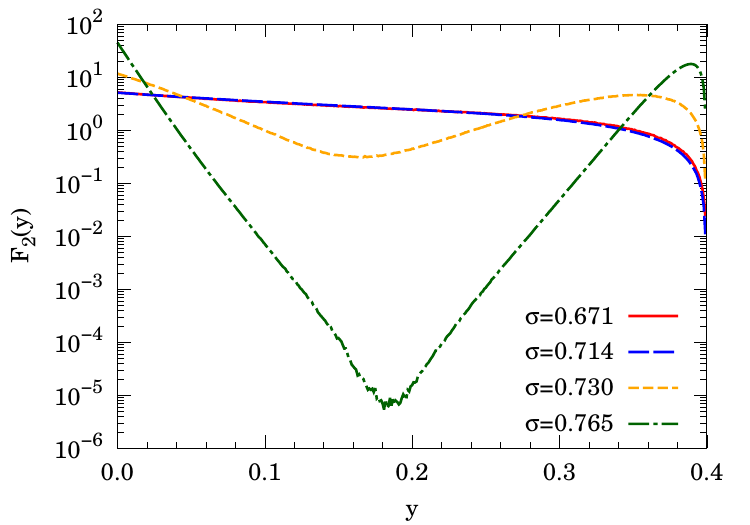}
\end{center}  
\caption{TDPP for $N=100$ HD with $\epsilon=0.4$ and 
\red{$\sigma=0.671,0.714,0.730,0.765$ corresponding to 
$\lambda=0.939,1.000,1.022,1.071$ respectively}. 
We can see how this density function starts changing only really near to the 
close packing density $\lambda_{\rm cp}=1.091$, when the TDP becomes very small 
at $y\approx 0$.}
\label{fig:fyy-HD}
\end{figure}

\subsection{Hard Disks (HD)}
We tried to reproduce the case $\epsilon=\sqrt{3}/2, \lambda=\sigma w=1$ of Fig. 
5(a) of Ref. \cite{Montero2023}. Our results for the LDF, LSF, and RDF are shown 
in Figs. \ref{fig:gx-HD}, \ref{fig:sk-HD}, and \ref{fig:gr-fig10} respectively. In
Fig. \ref{fig:snap-HD} we show a snapshot of the simulation box.

We also run simulations for the cases considered in Fig. 10 of Ref. 
\cite{Montero2023}. The results are shown in Fig. \ref{fig:gr-fig10}. Comparison 
with the work of Montero and Santos \cite{Montero2023} shows that our RDF 
is different from what they define as $g_{2D}$ \cite{Andres-appA}.

It is interesting to study how the solution of Montero and Santos 
\cite{Montero2023} performs outside of the nearest neighbor regime where it is 
expected to be not exact anymore. Such a study was carried out at the level 
of the compressibility factor in Fig. 7 of Ref. \cite{Montero2023a}. We want here 
to repeat it for the structure. In Fig. \ref{fig:gx-fig7} we show the comparison 
for the LDF between our exact MC simulations and the approximate solution of 
Montero and Santos for HD at $\lambda=1.2$ and $\epsilon=0.9, 1.0, 1.118$.
From the comparison we see that the solution of Montero \& Santos, which is exact 
for $\epsilon\le\epsilon_{\rm nn-HD}$, is a rather good approximation for 
$\epsilon_{\rm nn-HD}<\epsilon<\epsilon_{\rm sf}$, but it becomes a poor
approximation for $\epsilon\ge\epsilon_{\rm sf}$. The breakdown of their solution 
at $\epsilon>1$ manifests itself through an LDF that does not follow the 
exact result from the MC simulation. This confirms the findings of Kofke and Post 
\ref{Kofke1993}.

\red{It is interesting to note that Hu and Charbonneau \cite{Hu2021} has shown 
how the envelope of the LDF $g(x)-1$ has an exponential decay at large distances.}
  
\subsection{Square-Wells (SW) and Square-Shoulders (SS)}
For SW/SS we explored the following two limiting nearest neighbor cases 
considered in Table \ref{tab:tabI}, namely: 
(a) $\epsilon=4/5$, $r_0=6/5$ and (b) $\epsilon=\sqrt{7}/4$, $r_0=3/2$, with 
$|\beta\varphi_0|=1$, and a surface density $\sigma=6/10,7/10$.  Our results 
for the LDF, LSF, and RDF are shown in Figs.
\ref{fig:gx-SW}, \ref{fig:sk-SW}, and \ref{fig:gr-SW} respectively. In
Fig. \ref{fig:snap-SW} we show a snapshot of the simulation box for SW 
case (a) with $\sigma=7/10$.

Our results show how the two cases SW and SS have very similar structures in the
confined geometry under the nearest neighbor condition near close packing. The 
difference in structure between the two cases can be better seen at the level of 
the RDF where the SW produce a negative jump at $r=r_0$ whereas the SS produce a 
positive jump as expected.

\red{It would be an interesting project to explore how the {\sl sticky limit} 
is approached in this constrained geometry \cite{Fantoni05a,Fantoni05b,Fantoni06a,Fantoni06b,Fantoni06c,Fantoni07,Fantoni17c,Fantoni18b}.}

\section{Conclusions}
\label{sec:conclusions}

In this work we performed Monte Carlo computer experiments to extract meta data 
for the thermodynamic and structural properties of Hard-, Square-Well and 
Square-Shoulder disks in narrow channels. We worked in the canonical ensemble. Our 
data is subject only to the statistical (we never used more than $10^9$ single 
particle moves) and finite size errors (we used always 100 particles).

The novelty respect to previous studies relies in the use of the canonical ensemble
instead of the isothermal isobaric one and in the study of both the radial and
the linear distribution functions and of both the longitudinal and transverse 
pressures.

We compare our exact results \red{for Hard-Disks} with the semi-analytic ones 
of Montero and Santos \cite{Montero2023a,Montero2023} which are also exact in the 
nearest neighbor regime. We further compare our results with the results of the same 
authors but when the nearest neighbor condition is not met, making their solution 
just an approximation. In particular we see how such theoretical solution ceases to 
be a good approximation as soon as the single file condition is violated.

Regarding the comparison with the works of Montero and Santos, it is important to point 
out that the "exact" approach of those authors is based on a mapping to a pure 1D system, 
while our simulations deal with a true (confined) 2D system. Thus, our results 
reinforce the exact character of their method.

\red{We are aware that Montero and Santos are currently working at extending their 
theoretical framework to include the description of particles with a potential tail 
which would make possible the comparison with our Monte Carlo simulations of the 
Square-Well and Square-Shoulder particles.}

\begin{figure}[h!]
\begin{center}
\includegraphics[width=9cm]{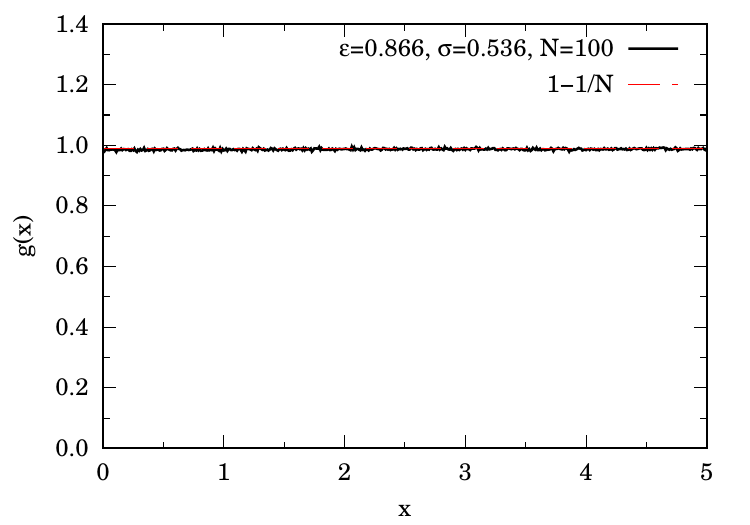}
\end{center}  
\caption{LDF for the ideal gas with $\lambda=1$ and $\epsilon=\sqrt{3}/2$.
The MC data are fitted very well by the exact result of 
$g_{\rm id}(x)=1-1/N$.}
\label{fig:gx-id}
\end{figure}
\begin{figure}[h!]
\begin{center}
\includegraphics[width=9cm]{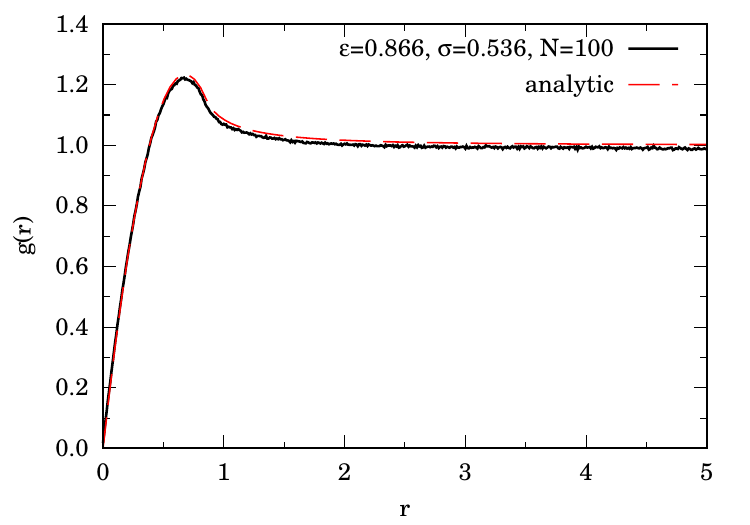}
\end{center}  
\caption{RDF for the ideal gas with $\lambda=1$ and $\epsilon=\sqrt{3}/2$. Here 
the $g(r)$ is calculated from Eq. (\ref{rdf}) using for $n_{\rm id}(r)$ its
asymptotic value $2\lambda\Delta$ everywhere. The analytic result is the one in
the thermodynamic limit of Eq. (\ref{eq:andres}). The slight discrepancy is the
expected finite size effect. Remember that $\lim_{r\to\infty}g_{\rm id}(r)=1-1/N$.}
\label{fig:gr-id}
\end{figure}
\begin{figure}[h!]
\begin{center}
\includegraphics[width=9cm]{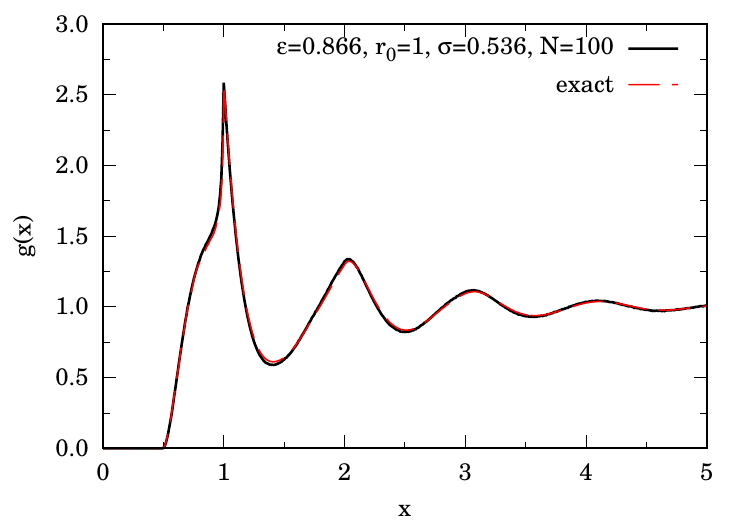}
\end{center}  
\caption{LDF for $N=100$ HD of radius $r_0=1$ with $\epsilon=\sqrt{3}/2$ and 
$\lambda=1$.
Our MC data is fitted very well by the exact result of Ref. \cite{Montero2023} which 
is in the thermodynamic limit.}
\label{fig:gx-HD}
\end{figure}
\begin{figure}[h!]
\begin{center}
\includegraphics[width=9cm]{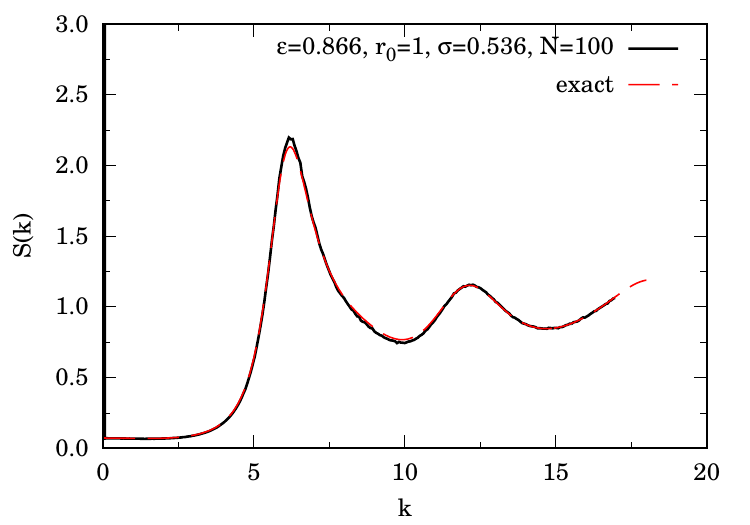}
\end{center}  
\caption{LSF for $N=100$ HD of radius $r_0=1$ with $\epsilon=\sqrt{3}/2$ and 
$\lambda=1$. We used $2n_{\rm max}+1$ wave numbers with $n_{\rm max}=270$.
The exact result in the thermodynamic limit, shared privately by A. Montero 
and not published before, fits our MC results very well.}
\label{fig:sk-HD}
\end{figure}
\begin{figure}[h!]
\begin{center}
\includegraphics[width=9cm]{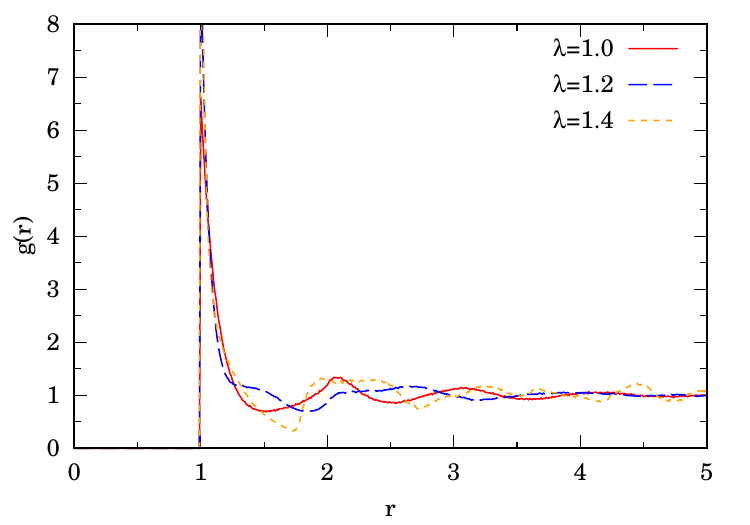}
\end{center}
\caption{RDF for $N=100$ HD of radius $r_0=1$ with $\epsilon=\sqrt{3}/2$ and 
$\lambda=1.0, 1.2, 1.4$. The contact value for the $\lambda=1.2,1.4$ cases is not
shown. To be compared with Fig. 10 of Ref. \cite{Montero2023}.}
\label{fig:gr-fig10}
\end{figure}
\begin{figure}[h!]
\begin{center}
\includegraphics[width=9cm]{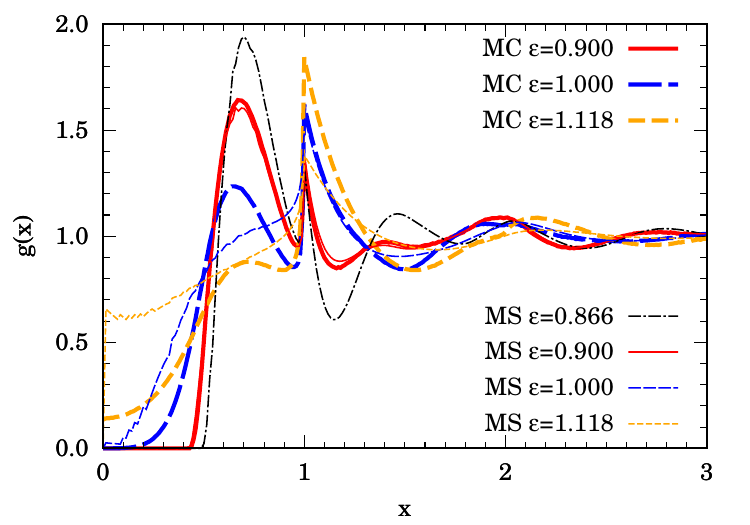}
\end{center}
\caption{LDF for $N=100$ HD of radius $r_0=1$ with $\lambda=1.2$ and 
$\epsilon=0.9, 1.0, 1.118$. Comparison between our exact MC simulation (thick
lines) and the theoretical approximate solution of Montero and Santos (MS) 
of Refs. \cite{Montero2023a,Montero2023} (thin lines). For the MS data we also 
show the exact result at $\epsilon=\sqrt{3}/2$ already published in Fig. 5(a) of 
Ref. \cite{Montero2023}. The remaining theoretical MS data was
shared privately by A. Montero and was not published before.}
\label{fig:gx-fig7}
\end{figure}
\begin{figure}[h!]
\begin{center}
\includegraphics[width=9cm]{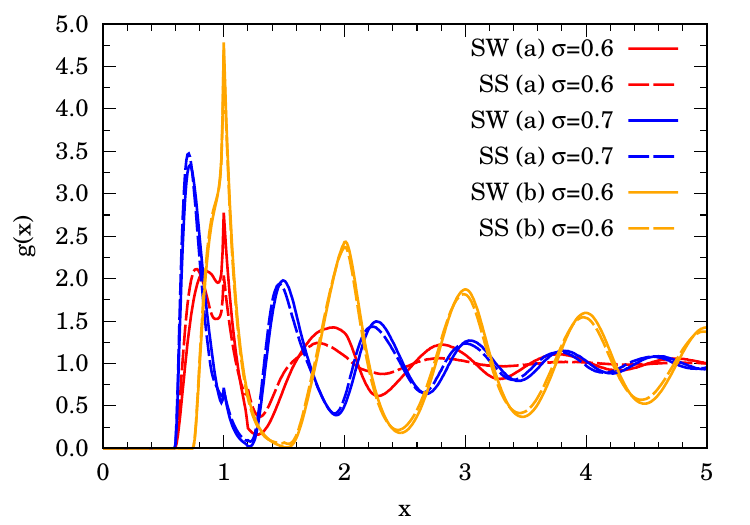}
\end{center}  
\caption{LDF for $N=100$ SW/SS cases (a) with $\sigma=6/10,7/10$ and (b) with 
$\sigma=6/10$.}
\label{fig:gx-SW}
\end{figure}
\begin{figure}[h!]
\begin{center}
\includegraphics[width=9cm]{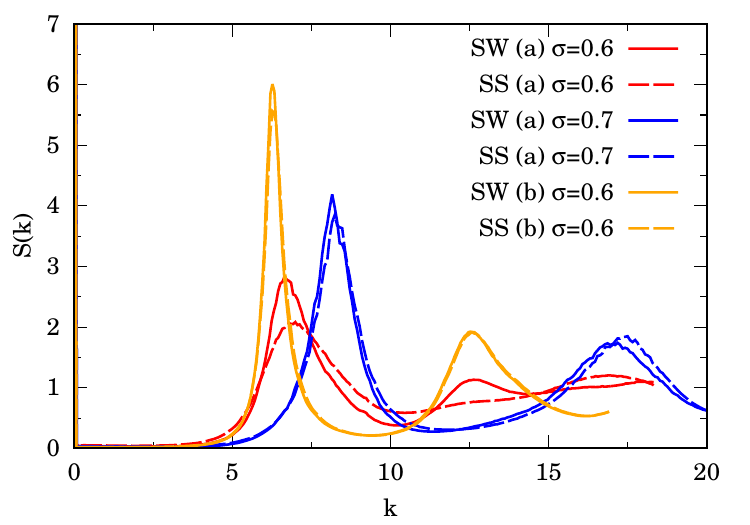}
\end{center}  
\caption{LSF for $N=100$ SW/SS cases (a) with $\sigma=6/10,7/10$ and (b) with 
$\sigma=6/10$. 
We used $2n_{\rm max}+1$ wave numbers with $n_{\rm max}=270$.}
\label{fig:sk-SW}
\end{figure}
\begin{figure}[h!]
\begin{center}
\includegraphics[width=9cm]{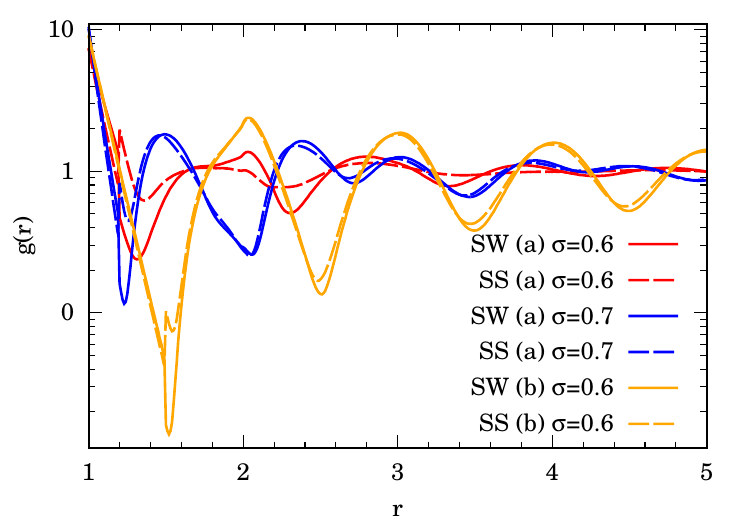}
\end{center}  
\caption{RDF for $N=100$ SW/SS cases (a) with $\sigma=6/10,7/10$ and (b) with 
$\sigma=6/10$. Clearly $g(r)=0$ for $r<1$. Note the logarithmic scale on the 
ordinates.}
\label{fig:gr-SW}
\end{figure}
%

\begin{acknowledgments}
I am grateful to Ana M. Montero and Andr\'es Santos for proposing the project,
stimulating its publication, and for the very many fruitful discussions and 
profound insights. I am grateful to Ana M. Montero for providing me with her 
results used in Figs. \ref{fig:tabI-HD}, \ref{fig:tdp-HD}, \ref{fig:gx-HD},
\ref{fig:sk-HD}, \ref{fig:gx-fig7} some of which had not been published before.
\end{acknowledgments}

\section*{Author declarations}
\subsection*{Conflict of interest}
The author has no conflicts to disclose.

\section*{Data availability}
The data that support the findings of this study are available from the 
corresponding author upon reasonable request.

\appendix

\section{On the longitudinal pressure of HD from the LDF}
\label{app:B}

Using the notation of Refs. \cite{Montero2023a,Montero2023} we have for the 
Equation Of State (EOS)
\bq \label{appb:1}
Z_L=\frac{\beta p}{\lambda}=
1+A^2\sum_{i,j}\phi_i\phi_ja_{ij}e^{-\beta p a_{ij}},
\eq
where $A^2$ and $\phi_i$ are the solutions to
\bq \label{appb:2}
\sum_je^{-\beta p a_{ij}}\phi_j=\frac{\beta p}{A^2}\phi_i.
\eq
The LDF in the range $a(\epsilon)<x<2a(\epsilon)$ is
\bq \label{appb:3}
g(x)=\frac{A^2}{\lambda}\sum_{i,j}\phi_i\phi_ja_{ij}e^{-\beta p x}
\Theta(x-a_{ij}).
\eq 
Our aim is to express the EOS in terms of the integrals
\bq \label{appb:4}
I_n=\int_{a(\epsilon)}^1dx\,x^ng(x),~~~~~n=0,1
\eq
Inserting Eq. (\ref{appb:3}) into Eq. (\ref{appb:4})
\bqs
\bq
I_0&=&\frac{A^2}{\beta p\lambda}\sum_{i,j}\phi_i\phi_j
\left(e^{-\beta p a_{ij}}-e^{-\beta p}\right),\\ \nonumber
I_1&=&\frac{A^2}{(\beta p)^2\lambda}\sum_{i,j}\phi_i\phi_j
\left[e^{-\beta p a_{ij}}(1+\beta p a_{ij})\right.\\
&&\left.-e^{-\beta p}(1+\beta p)\right].
\eq
\eqs
From Eq. (\ref{appb:2}) we have 
$\sum_{i,j}\phi_i\phi_je^{-\beta p a_{ij}}=\beta p/A^2$. Therefore
\bqs
\bq \label{eq:I0}
\lambda I_0&=&1-\frac{A^2}{\beta p}e^{-\beta p}\sum_{i,j}\phi_i\phi_j,\\
\nonumber
\beta p\lambda I_1&=&1+A^2\left[\sum_{i,j}\phi_i\phi_ja_{ij}e^{-\beta p a_{ij}}
\right.\\
\label{eq:I1}
&&\left.-e^{-\beta p}\left(1+\frac{1}{\beta p}\right)\sum_{i,j}\phi_i\phi_j\right].
\eq
\eqs
Comparison with Eq. (\ref{appb:1}) yields
\bq
\beta p\lambda I_1=Z_L-(1+\beta p)(1-\lambda I_0).
\eq
This is a linear equation in $Z_L$ which is solved by Eq. (\ref{eq:Z1DDD}) 
in the main text. From which immediately follows that for the pure 1D (Hard Rods)
case we find $Z_L=1/(1-\lambda)$, since $\epsilon\to 0$ and $a(\epsilon)\to 1$ so 
that $I_n=0$, as it should be \cite{Tonks1936}.

Note also that from Appendix C of Ref. \cite{Montero2023} follows that in
the $p\to\infty$ limit or equivalently in the $\lambda\to\lambda_{\rm cp}$ limit 
one finds
$\lim_{\lambda\to\lambda_{\rm cp}}\lambda I_0=\lim_{\lambda\to\lambda_{\rm cp}}\lambda^2 I_1=1$. 
In the continuum limit one has from Eq. (\ref{eq:I0})
\bqs
\bq
\lambda I_0&=&1-\frac{e^{-\beta p}}{\ell}J^2,\\
J&=&\int_{-\epsilon/2}^{\epsilon/2}\phi(y)\,dy.
\eq
\eqs
In the high pressure regime
\bqs
\bq
\phi(y)&\to&\frac{1}{\sqrt{\cal N}}[\phi_+(y)+\phi_-(y)],\\
\phi_{\pm}(y)&=&e^{-a(y\pm\epsilon/2)\beta p},\\
{\cal N}&\to&\frac{a(\epsilon)}{\epsilon\beta p}e^{-2a(\epsilon)\beta p},\\
\ell&\to&\frac{a(\epsilon)}{2\epsilon\beta p}e^{-a(\epsilon)\beta p}.
\eq
\eqs
Thus
\bq
J&=&\frac{2}{\sqrt{\cal N}}\int_{-\epsilon/2}^{\epsilon/2}\phi_+(y)\,dy.
\eq
By expanding $a(y+\epsilon/2)$ around $y=\epsilon/2$
\bq
a(y+\epsilon/2)\to a(\epsilon)+\frac{\epsilon}{a(\epsilon)}(\epsilon/2-y)
+\ldots.
\eq
Therefore
\bq \nonumber
J&\to&\frac{2}{\sqrt{\cal N}}e^{-a(\epsilon)\beta p}
\int_{-\epsilon/2}^{\epsilon/2}e^{-\frac{\epsilon\beta p}{a(\epsilon)}(\epsilon/2-y)}\,dy\\
&\to&\frac{2}{\sqrt{\cal N}}e^{-a(\epsilon)\beta p}
\frac{a(\epsilon)}{\epsilon\beta p}=2\sqrt{\frac{a(\epsilon)}{\epsilon\beta p}}. 
\eq
Consequently
\bq
\lambda I_0\to 1-8e^{-\beta p[1-a(\epsilon)]}.
\eq
Consistency between this result and Eq. (\ref{eq:hp}) 
gives Eqs. (\ref{eq:hdlim1})-(\ref{eq:hdlim2}) in the main text.

\section{RDF of the ideal gas in a narrow channel}
\label{app:A}

We arrive at the analytically exact Eq. (\ref{eq:andres}) for the RDF of the 
ideal gas confined in the narrow channel with the following steps
\bq \nonumber
g_{\rm id}(r)&=&\frac{\lambda}{2N}\int_{0}^{L}dx_1\int_{0}^{L}dx_2
\int_{-\epsilon/2}^{\epsilon/2}dy_1\,\frac{1}{\epsilon}
\int_{-\epsilon/2}^{\epsilon/2}dy_2\,
\frac{1}{\epsilon}\times\\
&&\delta(r-\sqrt{(x_2-x_1)^2+(y_2-y_1)^2}).
\eq
Since the integrand depends only on $x=|x_2-x_1|$ we have 
$\int_0^Ldx_1\int_0^Ldx_2\ldots=2\int_0^Ldx\,(L-x)\ldots$. Moreover
\bq
\delta(r-\sqrt{x^2+s^2})=\frac{r}{x}\delta(x-\sqrt{r^2-s^2}).
\eq
Therefore
\bq \nonumber
&&g_{\rm id}(r)=\\ \nonumber
&&\frac{\lambda}{N\epsilon^2}r\int_{-\epsilon/2}^{\epsilon/2}dy_1
\int_{-\epsilon/2}^{\epsilon/2}dy_2\,
\left(\frac{L}{\sqrt{r^2-(y_2-y_1)^2}}-1\right)=\\ \nonumber
&&\frac{2}{\epsilon^2}r\int_0^{\min(\epsilon,r)} ds\,(\epsilon-s)
\left(\frac{1}{\sqrt{r^2-s^2}}-\frac{1}{L}\right)\approx\\\label{eq:id-exact}
&&\frac{2}{\epsilon^2}r\int_0^{\min(\epsilon,r)} ds\,
\frac{\epsilon-s}{\sqrt{r^2-s^2}}.
\eq
Where in the first step we have assumed that $\sqrt{r^2-(y_2-y_1)^2}<L$ and 
in the third step we have taken the limit $L\to\infty$. In the limit 
$r\gg 1$, $\sqrt{r^2-s^2}\approx r$, so that $g_{\rm id}(r)\approx 1$ as 
expected.

The integral in Eq. (\ref{eq:id-exact}) can be analytically performed and the 
result is given by Eq. (\ref{eq:andres}) in the main text.


%

\end{document}